\documentclass[letterpaper,10pt,conference]{IEEEtran}
\usepackage[T1]{fontenc}
\usepackage[latin9]{inputenc}
\usepackage[cmex10]{amsmath}
\usepackage{amsthm}
\usepackage{graphicx}
\usepackage{epstopdf}
\usepackage{esint}
\usepackage{color}
\usepackage{wrapfig,psfrag}
\usepackage{amssymb}
\usepackage{setspace}
\usepackage{cite}
\usepackage{ifthen}
\makeatletter
\theoremstyle{plain}
\newtheorem{thm}{\protect\theoremname}
\theoremstyle{definition}
\newtheorem{defn}{\protect\definitionname}
\theoremstyle{remark}

\theoremstyle{plain}
\newtheorem{lem}{\protect\lemmaname}
\theoremstyle{definition}

\makeatother

\providecommand{\definitionname}{Definition}
\providecommand{\examplename}{Example}
\providecommand{\lemmaname}{Lemma}
\providecommand{\remarkname}{Remark}
\providecommand{\theoremname}{Theorem}

\newcommand{\Aseq}[1][1]{A_{#1}^{\infty}}
\newcommand{\aseq}[1][1]{a_{#1}^{\infty}}
\newcommand{\xseq}[1][1]{x_{#1}^{\infty}}

\newcommand{\yseq}[1][n]{y_{0}^{#1}}
\newcommand{\Dseq}[1][n]{D_0^{#1}}

\newcommand{\dseq}[1][n]{d_0^{#1}}

\newcommand{\nn}{\nonumber\\}
\newcommand{\bracket}[1]{\left[#1\right]}
\DeclareMathOperator{\Prob}{P}
\newcommand{\prob}[2][\relax]{\Prob_{#1}\bracket{#2}}

\DeclareMathOperator{\Eop}{E}
\providecommand{\E}[2][\relax]{\Eop_{#1}\bracket{#2}}

\newcommand{\limty}[1]{\lim_{#1\to\infty}}

\newcommand{\eqnref}[1]{(\ref{eqn:#1})}

\newcommand{\eqnlabel}[1]{\label{eqn:#1}}
\newcommand{\paren}[1]{\left(#1\right)}

\newcommand{\fe}[2][\relax]{%
\ifthenelse{\equal{#1}{\relax}}{f_{#2}}{f_{#2}(#1)}}

\newcommand{\fg}[2][\relax]{%
\ifthenelse{\equal{#1}{\relax}}{g_{#2}}{g_{#2}(#1)}}

\newcommand{\liminfp}[1][n\to\infty]{%
\liminf_{#1}\,\text{-p}\ }
\begin{document}

\title{Bits Through Bufferless Queues}


\author{\IEEEauthorblockN{Mehrnaz Tavan, Roy D.~Yates, and Waheed U.~Bajwa}
\IEEEauthorblockA{Department of Electrical and Computer Engineering\\
Rutgers University\\
Email: mt579@eden.rutgers.edu, ryates@winlab.rutgers.edu, waheed.bajwa@rutgers.edu}}

\maketitle
\begin{abstract}
This paper investigates the capacity of a channel in which information is
conveyed by the timing of consecutive packets passing through a
queue with independent and identically distributed service times. Such timing
channels are commonly studied under the assumption of a work-conserving
queue. In contrast, this paper studies the case of a bufferless queue that
drops arriving packets while a packet is in service. Under this
bufferless model, the paper provides upper bounds on the capacity of timing
channels and establishes achievable rates for the case of bufferless M/M/1
and M/G/1 queues. In particular, it is shown that a bufferless M/M/1 queue at
worst suffers less than 10\% reduction in capacity when compared to
an M/M/1 work-conserving queue.
%
%
\end{abstract}

\section{Introduction\label{sec:Introduction}}

Timing channels convey information by the
timing of consecutive packets -- rather than by their contents. Such channels
not only arise in many engineering contexts, such as covert communications
\cite{radosavljevic1992hiding,giles2002information} and sensor networks
\cite{morabito2011exploiting}, but can also provide a reasonable abstraction
of interactions in biological systems \cite{krishnaswamy30bacteria}. In
addition, information theoretic understanding of timing channels can
potentially help us attack the challenging problem of causal inference in
systems where causal relationships are determined by timing
information \cite{quinn2011estimating,liu2010information}. 

The study of information theoretic timing channels began in the seminal paper
\cite{anantharam1996bits}, which characterizes the capacity of a timing
channel described by a single-server timing queue (SSTQ) with independent and
identically distributed (iid) service times. In particular, we have from
\cite{anantharam1996bits} that the capacity of an SSTQ with iid
exponential service times ($\cdot$/M/1 queue) is equal to $e^{-1}$ nats per
average service time.

In this paper, we are also interested in studying the capacity of a timing
channel described by an SSTQ. However, in contrast to
\cite{anantharam1996bits}, our focus is on a \emph{bufferless} SSTQ that
discards incoming packets while a packet is in service. Bufferless SSTQs, despite their apparent
simplicity, are effective in mathematically modeling some systems
including protein synthesis networks \cite{osorio2012tractable}. Our interest in bufferless  SSTQs is related to the mutual information in tweet sequences. Suppose Bob receives tweets from Alice and occasionally tweets in response. While formulating a response, Bob ignores subsequent tweets from Alice. In this model, we can view Alice's tweets as arrivals and Bob's tweets as departures from a queue. The time Bob spends formulating a response is the service time of a tweet admitted to the system.  While the bufferless SSTQ is a simple model, it provides a starting point for characterizing how much information can be gleaned from  tweet timing data.

To the best of our knowledge, however, the capacity of
bufferless SSTQs in the context of timing channels has not been explored in
prior work. And while the bufferless SSTQ shares some similarities with the buffered SSTQ
in \cite{anantharam1996bits}, we will see that analyzing its capacity presents some new challenges in the absence of a one-to-one correspondence between
incoming and departing packets.

In this paper, we make the following contributions to the capacity analysis of timing channels described by bufferless SSTQs with iid service
times. First, we describe the maximum likelihood (ML) decoder for decoding
timing messages transmitted through a bufferless queue. Second, we provide a single-letter upper bound on the channel capacity under
arbitrary service distributions for the case of iid inter-arrival packet
times. Next, we provide a
single-letter upper bound and a looser closed-form upper bound on the channel
capacity under arbitrary service distributions.  Finally, we provide achievability results for bufferless M/M/1 and
M/G/1 queues using information density methods  \cite{verdu1994general}. In particular,
for the bufferless M/M/1 queue, achievable rates are shown to coincide with our outer bound. In addition, it is shown that a bufferless M/M/1 queue at
worst suffers less than 10\% reduction in achievable rate when compared to
an M/M/1 queue with infinite buffer \cite{anantharam1996bits}.

We conclude with a brief discussion of other related work on timing
channels. The setup studied in \cite{anantharam1996bits} corresponds to a
continuous timing channel. A discrete-time version of this timing channel is
analyzed in \cite{bedekar1998information,thomas1997shannon}. In
\cite{sundaresan2006capacity,coleman2009simple}, the authors revisited the
timing channel of \cite{anantharam1996bits} and provided capacity analysis
from the viewpoint of point processes. Finally, extensions of
\cite{anantharam1996bits} for the case when the distribution of service times
has bounded support is investigated in \cite{sellke2007capacity} and for the
case of a compound timing channel described by a tandem of queues is analyzed
in \cite{mimcilovic2006mismatch}. In all these works, however, the
fundamental assumption is that the queues are work conserving.

The rest of the paper is organized as follows. In
Section~\ref{sec:system-model}, we provide an overview of our system,
describe the optimal receiver, and provide a formal definition of capacity in
our setup. Section~\ref{sec:ConverseTheorem} derives outer bounds on the timing capacity that hold for all arrival processes. Section~\ref{sec:converse2} provides outer bounds for specific arrival processes and service time distributions. Section~\ref{sec:Achievability} investigates achievable rates in our system
and compares them to the outer  bounds obtained in Sections~\ref{sec:ConverseTheorem} and~\ref{sec:converse2}. Concluding remarks are in
Section~\ref{sec:conclusion}.

Note that we use $f_{X}(\cdot)$ to denote the probability density function (PDF) of random variable $X$. Similarly $f_{X|Y}(\cdot|\cdot)$ is the conditional PDF of $X$ given $Y$. In addition, $\exp(\cdot)$ denotes the inverse of $\log x$: $\exp(\log x)=x$.



\begin{figure}[t]
\centering
\psfrag{A1}{$\mathbf{A}_1$}
\psfrag{A2}{$\mathbf{A}_2$}
\psfrag{A3}{$\mathbf{A}_3$}
\psfrag{A4}{$\mathbf{A}_4$}
\psfrag{A5}{$\mathbf{A}_5$}
\psfrag{A6}{$\mathbf{A}_6$}
\psfrag{W1}{$\mathbf{S}_0$}
\psfrag{W2}{$\mathbf{W}_0$}
\psfrag{W3}{$\mathbf{W}_1$}
\psfrag{W4}{$\mathbf{W}_2$}
\psfrag{S2}{$\mathbf{S}_1$}
\psfrag{S3}{$\mathbf{S}_2$}
\psfrag{S4}{$\mathbf{S}_3$}
\psfrag{D0}{$\mathbf{D}_0$}
\psfrag{D1}{$\mathbf{D}_1$}
\psfrag{D2}{$\mathbf{D}_2$}
\psfrag{D3}{$\mathbf{D}_3$}

\includegraphics[scale=0.275]{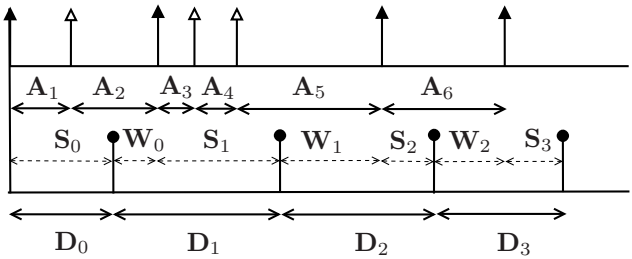}
\caption{\label{fig:System model} One realization of the input  and
 output sequence of the system is illustrated. The arrows with hollow arrowheads
show the packets arriving at the server that are dropped, the arrows
with solid arrowheads show the packets that enter the server,
and the lines with circles on top show the packets departing the
server.}
\end{figure}

\section{System Model \label{sec:system-model}}
The basic idea of the timing channel in \cite{anantharam1996bits}
is to use packet inter-arrival times to the server to encode a message. The receiver, based on
the departure times of packets from the server, decides which message has been
transmitted. In contrast to \cite{anantharam1996bits}, we consider a channel
that consists of a single-server bufferless queue with a zero packet waiting
room. Upon arrival at an idle server, a packet immediately enters service;
otherwise, if the server is busy with a previous packet, the incoming packet
is blocked and discarded.

In the following, we use $S_i$ to denote the service time of the $i^{th}$
packet admitted to service. As is customary in queuing systems, we assume
that service times are iid random variables,
independent of packet arrival times. Thus we refer to the timing channel induced by the (bufferless) queue with service time $S$ as the (bufferless) timing channel $S$. In Fig.~\ref{fig:System model}, one
realization of the input and output sequences, including arriving,
dropped  and departing packets, is illustrated under our setup. 



\subsection{Encoder}

The transmitted message is
represented by the discrete uniform index $U\in\left\{ 1,\cdots,M\right\} $. At the transmitter, each message $U=u$  will be encoded into an infinite sequence  of packet inter-arrival times  $\overline{A_{u}}=\left(A_0=0,A_{u,1},A_{u,2},\cdots\right)$
where $A_{u,j}$ is the inter-arrival time between packets $j-1$
and $j$  in codeword $u$. We refer to the packet submitted at time $A_0=0$ as packet zero. This packet carries no timing information and serves only to initialize the system. Similarly, we refer to packets $1,2,\ldots$ as codeword packets as their inter-arrival times define the codewords.
We note that using a codeword with infinite length is not a new phenomenon and
has been applied in \cite{visotsky2003optimum} to ARQ systems
where they design an infinite length codeword and transmit a part
of it to the receiver or in \cite{polyanskiy2011feedback} where an infinite
length codeword is transmitted and in the receiver, after observing
each packet, the decoding is performed.

\subsection{Decoder}

At the receiver, the decoder observes the inter-departure times
$D_{0},D_{1},\cdots,D_{n}$ where $D_0$ is the departure time of packet $0$
and $D_{i}$, $i>0$, denotes the time between packet departures $i-1$ and $i$.
These inter-departure times are used in estimating the index $V\in\left\{
1,\cdots,M\right\}$ corresponding to the transmitted message. A decoding
error occurs when $V\neq U$. In the bufferless queue, the subset of arrivals
that are admitted into service is denoted by the subsequence
$k_{0}=0,k_{1},\cdots$ such that
\begin{equation}
k_{i}=\min\left\{ m|\sum_{j=1}^{m}A_{j}-\sum_{j=0}^{i-1}D_{j}>0\right\} \eqnlabel{Tau_i}
\end{equation}
denotes the index of the  packet $i>0$ admitted to service. The
time that the server is idle between departure $i$ and the next arrival
is represented by $W_{i}$. Since the queue in our system is blocking
and has no buffer, the idling time $W_i$ can be represented as a deterministic function of the message index $U$ and prior departures $\Dseq[i]$ as
\begin{equation}
W_{i}(U,\Dseq[i])=\sum_{j=1}^{k_{i+1}}A_{U,j}-\sum_{j=0}^{i}D_{j}.\eqnlabel{W_i}
\end{equation}
For ease of notation, we use  $W_{i}(U,\Dseq[i])$ and the shorthand $W_i$ interchangeably. The
relationship between departure time $D_i$ and the corresponding
idling time and service time is
\begin{equation}
D_{i}=W_{i-1}(U,\Dseq[i-1])+S_{i}.\eqnlabel{D_i}
\end{equation}
Equivalent to  \eqnref{W_i} and \eqnref{D_i}, we can explicitly represent  $W_i$ and $D_i$ as functions of the arrival times $\Aseq$ and past departures $\Dseq[i-1]$:
\begin{align}
W_{i}(\Aseq,\Dseq[i])&=\sum_{j=1}^{k_{i+1}}A_{j}-\sum_{j=0}^{i}D_{j},\eqnlabel{W_i2}\\
D_{i}&=W_{i-1}(\Aseq,\Dseq[i-1])+S_{i}.\eqnlabel{D_i2}
\end{align}

After $n$ codeword packets are received, the MAP decoder
observes the departure times $\Dseq=\dseq$ and finds the most probable codeword
\begin{equation}
u^{*}(\dseq)=\arg\max_{u}P\left[U=u|\Dseq=\dseq\right]\eqnlabel{MAP}
\end{equation}
to have been transmitted.
Since the codewords are equiprobable, we can rewrite (\ref{eqn:MAP}) as the maximum likelihood problem
\begin{align}
 u^{*}(\dseq)\eqnlabel{My_MLE_1}
 &=\arg\max_{u}f_{\Dseq|U}\left[\dseq|u\right]\\
 &=\arg\max_{u}f_{D_{0}}[d_{0}]\prod_{i=1}^{n}f_{D_{i}|\Dseq[i-1],U}\left[d_{i}|\dseq[i-1],u\right].\eqnlabel{My_MLE_2}\\
&=\arg\max_{u}\sum_{i=1}^{n}\log f_{D_{i}|\Dseq[i-1],U}\left[d_{i}|\dseq[i-1],u\right].\eqnlabel{My_MLE_3}
 \end{align}
Since $W_{i-1}= W_{i-1}(U,\Dseq[i-1])$ is a deterministic function of $U,\Dseq[i-1]$,
\begin{align}
&f_{D_{i}|\Dseq[i-1],U}\left[d_{i}|\dseq[i-1],u\right]\nonumber\\
&\qquad=
f_{D_{i}|\Dseq[i-1],U,W_{i-1}}\left[d_{i}|\dseq[i-1],u,w_{i-1}\right]\eqnlabel{My_MLE_4}\\
&\qquad=
f_{S_{i}|\Dseq[i-1],U,W_{i-1}}\left[d_{i}-w_{i-1}|\dseq[i-1],u,w_{i-1}\right]\eqnlabel{My_MLE_5}\\
&\qquad=f_{S_{i}}\left[d_{i}-w_{i-1}(u,\dseq[i-1])\right].\eqnlabel{My_MLE_6}
\end{align}
Note that (\ref{eqn:My_MLE_5}) holds since  $D_{i}=w_{i-1}+S_{i}$ given $W_{i-1}=w_{i-1}$ and that (\ref{eqn:My_MLE_6}) follows since $S_{i}$ is independent of $U,\Dseq[i-1],W_{i-1}(U,\Dseq[i-1])$. Combining (\ref{eqn:My_MLE_3}) and (\ref{eqn:My_MLE_6}) and writing $w_{i-1}$ explicitly as a function of $u$ and $\dseq[i-1]$, we obtain
\begin{align}
 u^{*}(\dseq)=\arg\max_{u}\sum_{i=1}^{n}\log f_{S_{i}}\left[d_{i}-w_{i-1}(u,\dseq[i-1])\right].
  \eqnlabel{My_MLE_final}
\end{align}

\subsection{Capacity}
In this work, we aim to compute the  capacity  of the bufferless timing
channel. While each decoded message conveys $\log_{2}M$ bits of
information, the time required by the receiver to decode a message depends on
the packet departure times. In particular, we assume that the receiver
decodes after observing the departures of $n$ codeword packets. The expected
time required to observe these departures is
\begin{equation}
T_{n}=\sum_{i=0}^{n}E\left[D_{i}\right]=E\left[S_0\right]+\sum_{i=1}^{n}E\left[W_{i-1}+S_{i}\right].\eqnlabel{Time}
\end{equation}

Following \cite{anantharam1996bits,sundaresan2000robust} the achievable rate and the capacity for our system are defined as follows.
\begin{defn}
\label{Def:Capacity} If for every $\gamma>0$, a sequence of codewords from a
codebook with $M_{n}$ entries exists with
$\left(\log
M_{n}\right)/T_{n}>R-\gamma$ for all sufficiently large $n$, and the
corresponding maximum probability of error $\epsilon_n$ satisfying
$\lim_{n\rightarrow\infty}\epsilon_{n}=0$, then the rate $R$ is
achievable. The maximum
rate $R$ that satisfies this definition is called the
capacity of the timing channel and is denoted by $C$.  \end{defn}

%

\section{Converse Theorems\label{sec:ConverseTheorem}}
We follow the approach of \cite{anantharam1996bits} in deriving a converse. Using $P_e$ to denote the probability of a decoding error,  we observe that Fano's inequality \cite[sec.~2.10]{cover2012elements} and equiprobable $U$ imply
\begin{align}
H(U|V) & \leq H(P_e)+P_e\log M_{n}\eqnlabel{Fano_1}\\
 & \leq H(P_e)+\epsilon_{n}\log M_{n}\eqnlabel{Fano_2}\\
 & \leq\log2+\epsilon_{n}\log M_{n}\eqnlabel{Fano_3}\\
 & =\log2+\epsilon_{n}\log M_{n}+H(U)-\log M_{n},\eqnlabel{Fano_5}
\end{align}
where we assume that $H(P_e)\leq \log2$.
We can conclude
that
\begin{align}
\log M_{n} & \leq\frac{1}{1-\epsilon_{n}}\left[I(U;V)+\log2\right]\eqnlabel{Fano_6}\\
 & \leq\frac{1}{1-\epsilon_{n}}\left[I(\Aseq;\Dseq)+\log2\right],
 \eqnlabel{Data_Proc}
\end{align}
where (\ref{eqn:Data_Proc}) follows from
the data processing lemma \cite[sec.~2.8]{cover2012elements}.


Before stating a converse for our system, we need the following
lemmas.
\begin{lem}
\label{lem:IWS-equal-bound}
The mutual
information between the input codeword and the output departure times
satisfies
\begin{align}
I(\Aseq;\Dseq) & =\sum_{i=1}^{n}\left(h(W_{i-1}+S_{i}|\Dseq[i-1])-h(S_{i})\right).\eqnlabel{IWS-equal}
\end{align}
\end{lem}
\begin{IEEEproof}
By the chain rule,
\begin{equation}
I(\Aseq;\Dseq)
=I(\Aseq;D_{0})+\sum_{i=1}^{n}I(\Aseq;D_{i}|\Dseq[i-1]).
\eqnlabel{chain}
\end{equation}
Since $D_0=S_0$, which is independent of the code packet arrivals $\Aseq$,
\begin{equation}
I(\Aseq;D_{0})= I(\Aseq;S_{0})=0.\eqnlabel{StarA_1}
\end{equation}
Moreover,
\begin{align}
&I(\Aseq;D_{i}|\Dseq[i-1])\nonumber\\
&=
h(D_{i}|\Dseq[i-1])-h(D_{i}|\Aseq,\Dseq[i-1])\\
&=
h(W_{i-1}+S_{i}|\Dseq[i-1])
-h(W_{i-1}+S_{i}|\Aseq,\Dseq[i-1])\eqnlabel{Mut_InfA_1}\\
&=h(W_{i-1}+S_{i}|\Dseq[i-1])
-h(S_{i}|\Aseq,\Dseq[i-1],W_{i-1})\eqnlabel{Mut_InfA_3}\\
&=h(W_{i-1}+S_{i}|\Dseq[i-1])-h(S_{i})\eqnlabel{Mut_InfA_4}.
\end{align}
Note that (\ref{eqn:Mut_InfA_3}) holds
since $\Aseq,\Dseq[i-1]$ deterministically specify $W_{i-1}$
using \eqnref{W_i2}; (\ref{eqn:Mut_InfA_4}) holds since $S_{i}$ is independent of the arrivals $\Aseq$, the prior departures $\Dseq[i-1]$ and the idle period $W_{i-1}$. The lemma follows from (\ref{eqn:chain}), (\ref{eqn:StarA_1}) and (\ref{eqn:Mut_InfA_4}).
\end{IEEEproof}

\begin{lem}
\label{lem:IWS-outer-bound}
The mutual
information between the input codeword and the output departure times
satisfies
\begin{align}
I(\Aseq;\Dseq) & \leq\sum_{i=1}^{n}I(W_{i-1};W_{i-1}+S_{i}).\eqnlabel{IWS-outer}
\end{align}
\end{lem}
\begin{IEEEproof}
Based on Lemma~\ref{lem:IWS-equal-bound},
\begin{align}
I(\Aseq;\Dseq)
&=
\sum_{i=1}^{n}\left(h(W_{i-1}+S_{i}|\Dseq[i-1])-h(S_{i})\right)\nn
 & \leq \sum_{i=1}^{n}\left(h(W_{i-1}+S_{i})-h(S_{i})\right).\eqnlabel{Mut_InfA_5}
\end{align}
Note that (\ref{eqn:Mut_InfA_5}) holds
since conditioning reduces entropy. 
\end{IEEEproof}

To develop universal bounds valid for all arrival and service processes, we follow the approach in \cite{anantharam1996bits}  and define
\begin{equation}
c(a)\equiv\sup_{\substack{
X\geq0\\
E\left[X\right]\leq a}}I(X;X+S)\eqnlabel{concave}
\end{equation}
where $X$ is independent of $S$. We note that $c(a)$ is a monotone concave function
in the argument $a$, and that this will provide a universal upper bound on the
capacity of the timing channel. We start with a relaxation of
Lemma~\ref{lem:IWS-outer-bound}.

\begin{lem}
\label{lem:Defining-where-} The mutual information between the input codeword
and the output departure times satisfies
\begin{equation}
I(\Aseq;\Dseq)\leq\sum_{i=1}^{n}c\left(\E{W_{i-1}}\right).
\eqnlabel{Verdu_Ineq}
\end{equation}
\end{lem}

\begin{IEEEproof}
 Lemma~\ref{lem:IWS-outer-bound} and  (\ref{eqn:concave}) imply
%
\begin{align}
I(\Aseq;\Dseq) & \leq\sum_{i=1}^{n}I(W_{i-1};W_{i-1}+S_{i})\\
 & \leq\sum_{i=1}^{n}\sup_{\substack{
X_{i}\geq0\\
\E{X_{i}}\leq \E{W_{i-1}}
}}
I(X_{i};X_{i}+S_{i})\\
&=\sum_{i=1}^{n}c\left(\E{W_{i-1}}\right).
\end{align}
\end{IEEEproof}
Now using Lemma~\ref{lem:Defining-where-} , we can define a general
converse which is parallel to \cite[Thm.~2]{anantharam1996bits}.

\begin{thm} \label{thm:The-converse-theorem with c(a):} The timing channel $S$ with $\E{S}=1/\mu$ has capacity 
 \begin{align}
C\leq\overline{C}(S) & \equiv \sup_{\lambda>0}\frac{c(\frac{1}{\lambda})}{\frac{1}{\lambda}+\frac{1}{\mu}}.
\eqnlabel{Converse_1}
\end{align}
\end{thm}
\begin{IEEEproof}
Let 
\begin{equation}\eqnlabel{Rn-defn}
R_n=\frac{(1-\epsilon_n)\log M_n}{T_n}.
\end{equation} Combining  (\ref{eqn:Time}), (\ref{eqn:Data_Proc}), and Lemma~\ref{lem:Defining-where-} yields
\begin{align}
R_n
&\leq\frac{\frac{1}{n}\left[\sum_{i=1}^{n}c(E\left[W_{i-1}\right])+\log2\right]}{\frac{\E{S_0}}{n}+\frac{1}{n}\sum_{i=1}^{n}\left(\E{W_{i-1}}+\E{S_{i}}\right)}.\eqnlabel{Up_Bound_FD_4}
\end{align}
Defining $\lambda_n^{-1}=\frac{1}{n}\sum_{i=1}^{n}\E{W_{i-1}}$, concavity of $c(a)$ implies
\begin{align}
R_n&\leq\frac{c(\frac{1}{\lambda_n})+\frac{\log2}{n}}{\frac{1}{\lambda_n} +\frac{1}{\mu}}
\le \sup_{\lambda>0}\frac{c(\frac{1}{\lambda})}{\frac{1}{\lambda}+\frac{1}{\mu}} +\frac{\log 2}{n/\mu}.\eqnlabel{Up_Bound_FD_5}
\end{align}
The claim follows as  $n\rightarrow\infty$.
\end{IEEEproof}

We can further loosen Theorem~\ref{thm:The-converse-theorem with c(a):} by
making use of the following lemma.



\begin{lem}\label{lem:Upper_c}
For a timing channel $S$,  $c(a)$ defined in (\ref{eqn:concave}) satisfies
%
\begin{equation}
c(a)\leq \log(e)+\log(a+\E{S})-h(S).\eqnlabel{Upper_c}
\end{equation}
\end{lem}
\begin{IEEEproof}
Based on (\ref{eqn:concave}), we have
\begin{equation}
c(a)=\sup_{\substack{
E\left[X\right]\leq a\\
X\geq0
}}h(X+S)-h(S).\eqnlabel{Up_Bound_2}
\end{equation}
Notice that $h(X+S)$ subject to the constraints $E\left[X\right]\leq\ a$ and
$X\geq0$ and fixed service distribution will be maximized when $X+S$ has
exponential distribution with rate $\left(a+\E{S}\right)^{-1}$ \cite{cover2012elements}. The proof now
follows from the entropy of an exponential distribution.
\end{IEEEproof}
 However, there is no guarantee that for any given
service distribution, there exists a nonnegative random variable with
$\E{X}\leq a$ such that its summation with $S$ has exponential
distribution. As a result, $\log(e)+\log(a+\E{S})-h(S)$ is an upper bound on $c(a)$. A universal upper bound on the capacity of the system can now be stated.
\begin{thm}\label{thm:twopartUpperBound}
The bufferless timing queue $S$ with $\E{S}=1/\mu$ has capacity
\begin{align}
C & \leq\begin{cases}
\frac{\log e+\log\left(\frac{1}{\mu}\right)-h(S)}{\mu^{-1}}, & h(S)<\log\left(1/\mu\right),\\
\frac{\log e}{\exp\left(h(S)\right)}, & h(S)\geq\log\left(1/\mu\right).
\end{cases}
\end{align}
\end{thm}
\begin{IEEEproof}
Based on Theorem~\ref{thm:The-converse-theorem with c(a):} and Lemma~\ref{lem:Upper_c}, $R_n$ defined in \eqnref{Rn-defn}
satisfies
\begin{align}
R_n &\leq\sup_{\lambda>0}\frac{\log e+\log\left(\frac{1}{\lambda}+\frac{1}{\mu}\right)-h(S)}{\frac{1}{\lambda}+\frac{1}{\mu}}.\eqnlabel{Up_Bound_3}
\end{align}
By taking the derivative of the upper bound in (\ref{eqn:Up_Bound_3}) with respect to $\lambda^{-1}$, the optimal $\lambda$ will satisfy
\begin{equation}
h(S)=\log\paren{\frac{1}{\lambda^*}+\frac{1}{\mu}}.
\end{equation}
Since $\lambda$ is a nonnegative number, when $h(S)<\log(1/\mu)$, the supremum is approached as $\lambda^{-1}\rightarrow0$ and the universal upper bound will be
\begin{align}
R_n& \leq\mu\bracket{\log e+\log(1/\mu)-h(S)}.
\end{align}
Otherwise,
\begin{align}
R_n&\leq\frac{\log e}{\exp\left(h(S)\right)}.
\end{align}
\end{IEEEproof}

\section{Queue-specific Outer Bounds}\label{sec:converse2}
We note that Lemmas~\ref{lem:IWS-equal-bound} and \ref{lem:IWS-outer-bound} make no particular assumptions regarding the statistical structure of the arrivals. However, in the absence of such assumptions, memory in the arrivals can induce idling times $W_i$ that are difficult to characterize.
To go further, we focus on the special case of codebooks with iid inter-arrival times.
With iid inter-arrivals, each time a packet
enters service, the queue undergoes a renewal.
In particular, the $i$th renewal point marks the beginning of a service time $S_i$ and a set of subsequent iid packet inter-arrival times $A_{k_{i}+1},A_{k_i+2},\ldots$
such that the distributions of $S_i$ and $\left\{A_{k_i+j}\right\}$ are sufficient to evaluate the distribution of the number of packet arrivals that are dropped during the service as well as the idling time $W_i$ that follows the service completion. Because service times and inter-arrival times are both iid, a renewal occurs at the end of the idling period when the next arrival is admitted.  We note that $W_i$ depends on $S_i$; however the renewal implies that $\left(S_{0},W_{0}\right),\left(S_{1},W_{1}\right),\cdots,\left(S_{n},W_{n}\right)$
constitute independent tuples. This observation yields the following outer bound for iid inter-arrivals.
\begin{thm}\label{thm:CASbound}
With iid inter-arrival times identical to $A$, the bufferless timing channel $S$ has capacity $C$
satisfying
\begin{equation}
C \le \overline{C}(A,S) \equiv \frac{I(W;W+S)}{E\left[W\right]+E\left[S\right]},\eqnlabel{iidConverse_1}
\end{equation}
where $W$ is independent of $S$ but has the idling time distribution induced by $A$ and $S$.
\end{thm}
\begin{IEEEproof}
Since each service initiation marks a renewal, Lemma~\ref{lem:IWS-outer-bound} reduces to
\begin{align}
I(\Aseq;\Dseq) &\leq nI(W_{i-1};W_{i-1}+S_{i}).\eqnlabel{IWS-iid}
\end{align}
In addition, (\ref{eqn:Time}) yields
\begin{equation}
T_{n}=E\left[S_0\right]+n\left(E\left[W_{i-1}\right]+E\left[S_{i}\right]\right).\eqnlabel{Time-iid}
\end{equation}
Combining (\ref{eqn:Data_Proc}), (\ref{eqn:IWS-iid}) and (\ref{eqn:Time-iid}) yields
\begin{align}
R_n&\leq
\frac{n I(W_{i-1};W_{i-1}+S_{i})+\log2}{ %
\E{S_0}+n\left(\E{W_{i-1}}+\E{S_{i}}\right)}.
\eqnlabel{Main_Up_Bound_1}
\end{align}
The claim follows as $n\rightarrow\infty$.
 \end{IEEEproof}
In general, computing the PDF of $W$ is nontrivial as it can involve $n$-fold convolutions of the PDF of $A_i$.  Thus, the primary use of Theorem~\ref{thm:CASbound} is the case when the $A_i$ form a rate $\lambda$ Poisson arrival process. In this case, the idling times $W_i$ are exponential $(\lambda)$ random variables independent of $S$, and the queueing system is an M/G/1 single server bufferless queue. For Poisson arrivals, the outer bound $\overline{C}(A,S)$ reduces to a straightforward numerical evaluation of $I(W;W+S)$.

As a special case, we analyze the M/M/1 queue in which the service
time is exponential with rate $\mu$. In this case, $S$ will have entropy
\begin{equation}\eqnlabel{Hexpo}
h(S) = \log e + \log\frac{1}{\mu}
\end{equation}
and $D=W+S$ will have the hypoexponential
distribution
\begin{equation}\eqnlabel{hypo-pdf}
f_{D}(d)=\frac{\mu\lambda}{\mu-\lambda}\left(e^{-\lambda d}-e^{-\mu d}\right),\qquad d\geq0,
\end{equation}
and entropy $h(D)=h_{\text{hypo}}(\lambda,\mu)$. Since $I(W;W+S)=h(D)-h(S)$, Theorem~\ref{thm:CASbound} yields the outer bound
\begin{align}
\overline{C}(A,S)=R(\lambda,\mu)\eqnlabel{MM1Upper}
\end{align}
where 
\begin{align}
R(\lambda,\mu)
\equiv \frac{h_{\text{hypo}}(\lambda,\mu)-\log e+\log\mu}{1/\lambda+ 1/\mu}.\eqnlabel{Rdefn}
\end{align}
The entropy  $h_{\text{hypo}}(\lambda,\mu)$ cannot be computed in a closed form. Using numerical
integration methods, the  upper bound in (\ref{eqn:MM1Upper}) is computed as a function of
$\lambda/\mu$  as shown in Fig.~\ref{fig:Upperbound-ExpIdle} (see Appendix \ref{appendix:a} for proof that  \eqnref{Rdefn} is a function of $\lambda/\mu$ for fixed $\mu$). It can be seen
from this figure that when $\lambda/\mu$ is close to zero, corresponding to a
queue that is idle most of the time, the upper bound on capacity is also
close to zero; this is to be expected since the time required to receive $n$
packets will be large in this case. On the other hand, when $\lambda \gg \mu$, the expected idling
time reduces, but more and more packets are dropped, and it becomes difficult for
the receiver to decode messages, resulting in a decreasing upper bound on
capacity.


\begin{figure}
\psfrag{x2}{\hspace{3mm}$\mathbf{\lambda/\mu}$}
\psfrag{y2}{Rate (nats/$\E{S}$)}
\includegraphics[scale=0.55]{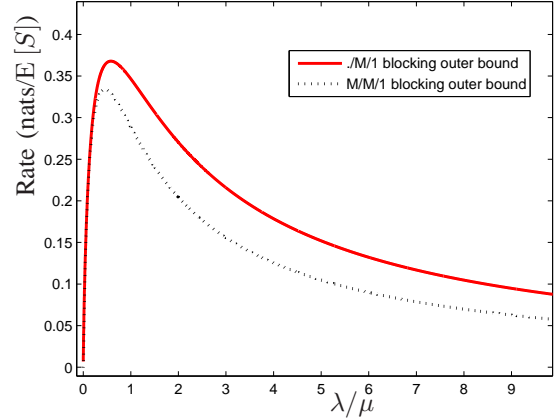}
\caption{\label{fig:Upperbound-ExpIdle} Comparison between the bufferless $\cdot$/M/1 and M/M/1 queue upper bounds for $0<\lambda/\mu<10$ where $\lambda$ is the arrival rate  and $\mu$ is the service rate. The M/M/1 upper bound coincides with the achievable rate for M/M/1 as well.}
\end{figure}


Fig.~\ref{fig:Upperbound-ExpIdle} compares the  Theorem~\ref{thm:twopartUpperBound} universal upper bound for the $\cdot$/M/1 queue to the upper bound derived for M/M/1 queue in (\ref{eqn:MM1Upper}). It can be seen from this figure that although the Theorem~\ref{thm:twopartUpperBound} bound is looser than (\ref{eqn:MM1Upper}), the two upper bounds almost coincide for $0<\lambda/\mu<0.2$.


\section{Achievability\label{sec:Achievability}}

While the upper bounds in Sections~\ref{sec:ConverseTheorem} and~\ref{sec:converse2} make use of the maximization of the mutual information between idling time and inter-departure time, the only parameters in our control for coding purposes are the inter-arrival times. In order for our system to achieve the Theorem~\ref{thm:twopartUpperBound} upper bound, two conditions  must be fulfilled: 1) The inter-departure times must be iid so $h(W_{k-1}+S_{k}|D_{1}^{k-1})=h(W_{k-1}+S_{k})$ which leads to equality in (\ref{eqn:IWS-outer}); 2) The inter-arrival times must be distributed such that asymptotically, the induced idling time maximizes (\ref{eqn:Converse_1}). The first condition is satisfied only when the service time is exponential; Otherwise, the relationship between $W_{i}$ and $S_i$ would create dependency between consecutive inter-departure times $D_i$ and $D_{i+1}$. When $S$ has exponential distribution, \cite[Theorem~3]{anantharam1996bits} shows that among the distributions with $\E{W+S}\leq 1/\lambda+1/\mu$,  $I(W;W+S)$ is maximized  when the distribution of  $W$ is a mixture of an exponential  with expected value $\mu^{-1}+\lambda^{-1}$ and an impulse at the origin. The resulting distribution for inter-departure time will be exponential with expected value $\mu^{-1}+\lambda^{-1}$ which is the distribution used in Theorem~\ref{thm:twopartUpperBound}. In our system, since $W$ cannot have zero value, the above conditions cannot be satisfied simultaneously and the Theorem~\ref{thm:twopartUpperBound} upper bound is not achievable.

\subsection{Achievability for the M/M/1 Queue}
To derive  achievability results, we use the information density method introduced in
\cite{verdu1994general}. For the bufferless timing queue, the information density is given by
 \begin{equation}
 i_{\Aseq;\Dseq}(\Aseq;\Dseq)=\log\frac{f_{\Dseq|\Aseq}\left(\Dseq|\Aseq\right)}{f_{\Dseq}\left(\Dseq\right)}.\eqnlabel{Inf_Density}
 \end{equation}
We will employ the following definition and theorem.
\begin{defn} The {\em liminf in probability} of a  sequence
of random variables  $Q_{n}$ is  
\begin{align*}
& \liminfp Q_n\\
& =\sup\left\{ \alpha>0|\lim_{n\rightarrow\infty}P\left[Q_{n}\leq\alpha-\gamma\right]=0,\forall \gamma>0\right\}.\eqnlabel{LimInfInProb}
\end{align*} 
\end{defn}
\begin{lem}[\cite{verdu1994general}]
\label{lem:-A-sufficient}
A sufficient condition
for rate $R$ to be achievable is existence of some input process
$\Aseq$ for which 
\begin{equation*}
\liminfp \left[\frac{1}{T_n}i_{\Aseq;\Dseq}(\Aseq;\Dseq)\right]\geq R.
\end{equation*}
\end{lem}
We will use Lemma~\ref{lem:-A-sufficient} to prove the following achievability result expressed in terms of $R(\lambda,\mu)$ given in \eqnref{Rdefn}. 
\begin{thm}
\label{thm:Achievab_MM1}The M/M/1 bufferless queue with service
rate $\mu$ and arrival rate $\lambda$ has capacity
\begin{equation*}
C(\lambda,\mu)\geq R(\lambda,\mu).
\eqnlabel{Achievab_MM1}
\end{equation*}
\end{thm}
\begin{IEEEproof}
In the M/M/1 queue, the arrival process is Poisson with rate
$\lambda$. As noted at the start of Section~\ref{sec:converse2}, the queue has a renewal each time a packet enters service. These inter-renewal
times are of the form $S_{i}+W_{i}$ where  $S_{i}$ and $W_{i}$ may be dependent, but $S_{i},W_{i}$ are independent of $S_{j},W_{j}$
for $j\neq i$. For Poisson arrivals, the memorylessness
of the exponential distribution implies $S_{i}$ and $W_{i}$ are independent. As a result, the inter-departure times $D_i$ are iid hypoexponential random variables with PDF given by  \eqnref{Hexpo}. Hence we can write
\begin{equation}
f_{\Dseq}(\dseq)=f_{S_{0}}(d_0)\prod_{i=1}^{n}f_{D_{i}}(d_{i}).\eqnlabel{DeparturePDF}
\end{equation}
 It follows from Lemma~\ref{lem:IWS-equal-bound} that the expected value of the information density will be
\begin{align}
\E{i_{\Aseq;\Dseq}(\Aseq;\Dseq)}
&=I(\Aseq;\Dseq)\\
 & =\sum_{k=0}^{n}\left[h(W_{k-1}+S_{k})-h(S_{k})\right]\eqnlabel{Lower-MM1}\\
 & =n\left(h(W+S)-h(S)\right)\eqnlabel{Lower-MM2}\\
 & =n\left(h_{\text{hypo}}(\lambda,\mu)-\log e+\log\mu\right).\eqnlabel{Lower-MM3}
\end{align}
Furthermore,
\begin{align}
 & f_{\Dseq|\Aseq}\left(\dseq|\aseq\right)\eqnlabel{Conditional departure_1}\nonumber\\
 & =f_{S_{0}}(d_0)\prod_{i=1}^{n}f_{D_{i}|\Aseq,\Dseq[i-1]}\left(d_{i}|\aseq,\dseq[i-1]\right)\\
 & =f_{S_{0}}(d_0)\prod_{i=1}^{n}f_{D_{i}|\Aseq,\Dseq[i-1],W_{i-1}}\left(d_{i}|\aseq,\dseq[i-1],w_{i-1}\right)\eqnlabel{Conditional departure_2}  \\
 & =f_{S_{0}}(d_0)\prod_{i=1}^{n}f_{S_i|\Aseq,\Dseq[i-1],W_{i-1}}\left(d_i-w_{i-1}|\aseq,\dseq[i-1],w_{i-1}\right)\eqnlabel{Conditional departure_3}\\
 & =f_{S_{0}}(d_0)\prod_{i=1}^{n}f_{S_{i}}\left(d_i-w_{i-1}(\aseq,\dseq[i-1])\right),\eqnlabel{Conditional departure_4}
\end{align}
where (\ref{eqn:Conditional departure_2}) holds
due to \eqnref{W_i2}, and (\ref{eqn:Conditional departure_3}) follows
due to \eqnref{D_i2}. Since
the server processes the packets independent of the arrival process, $S_{i}$ is independent of $\Aseq,\Dseq[i-1],W_{i-1}$,
and thus (\ref{eqn:Conditional departure_4}) holds.
Using (\ref{eqn:DeparturePDF}) and (\ref{eqn:Conditional departure_4}),
(\ref{eqn:Inf_Density}) normalized by $T_n$ can be written as
\begin{align}
 &\frac{1}{T_n} i_{\Aseq;\Dseq}(\Aseq;\Dseq)\nonumber\\
 &=\frac{1}{T_n}
 \left[\sum_{i=1}^{n}\log\left(f_{S_{i}}(D_i-W_{i-1})\right)-\sum_{i=1}^{n}\log\left(f_{D_{i}}(D_i)\right)\right]\nn
 &=\frac{n}{T_n}\frac{1}{n}
  \left[\sum_{i=1}^{n}\log\left(f_{S_{i}}(S_i)\right)-\sum_{i=1}^{n}\log\left(f_{D_{i}}(D_i)\right)\right],
  \eqnlabel{Inf_Den_5}
\end{align}
since $S_i=D_i-W_{i-1}$.  Since the $W_i$ are iid exponential $(\lambda)$ random variables, \eqnref{Time-iid} implies
\begin{equation}
\limty{n}\frac{n}{T_n}=\frac{1}{\E{W}+\E{S}}=\frac{1}{1/\lambda +1/\mu}.\eqnlabel{Tn-limit}
\end{equation}
By the strong law of large numbers \cite{ross1983stochastic}, it follows from \eqnref{Inf_Den_5} and \eqnref{Tn-limit} that
\begin{align}
\limty{n}\frac{i_{\Aseq;\Dseq}(\Aseq;\Dseq)}{T_n} 
&= \frac{h(D)-h(S)}{1/\lambda +1/\mu}
=R(\lambda,\mu)\quad\text{wp~1}.\eqnlabel{wp1limit}
\nonumber
\end{align}
It follows that the liminf in probability of $i_{\Aseq;\Dseq}(\Aseq;\Dseq)/T_n$ equals $R(\lambda,\mu)$ and thus by Lemma~\ref{lem:-A-sufficient}, rate $R(\lambda,\mu)$ is achievable.
\end{IEEEproof}

Comparing Theorem~\ref{thm:Achievab_MM1} and the upper bound (\ref{eqn:MM1Upper}), we see that the achievable rate $R(\lambda,\mu)$ matches the upper bound for the M/M/1 bufferless queue. Thus $R(\lambda,\mu)$ is the capacity of the bufferless M/M/1 timing channel with arrival rate $\lambda$ and service rate $\mu$. This M/M/1 capacity is illustrated in Fig.~\ref{fig:Upperbound-ExpIdle}. The maximum achievable rate in (\ref{eqn:MM1Upper}) is 0.3340 nats per average server time, and the maximum of  universal upper bound is 0.3679 which implies that a bufferless M/M/1 queue at worst suffers less than  $10\%$
reduction in achievable rate when compared to the universal upper bound.

\begin{figure}
\psfrag{x1}{\hspace{3mm}$\mathbf{\lambda/\mu}$}
\psfrag{y1}{Rate(nats/$\E{S}$)}
\includegraphics[scale=0.6]{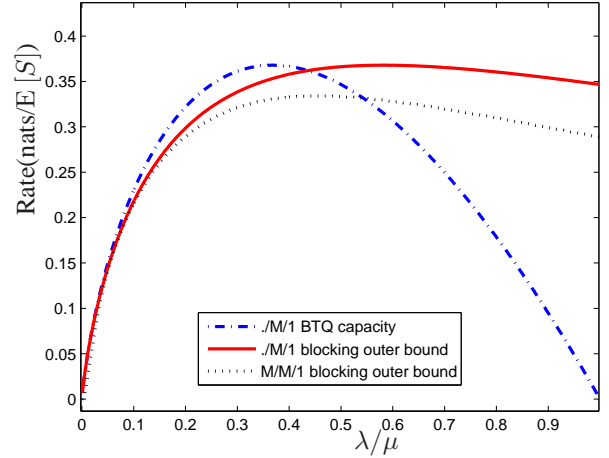}
\caption{\label{fig:Upperbound-Comp} Comparison of  $C(\lambda)$ of the "bits through queues" (BTQ) paper \cite[Theorem~4]{anantharam1996bits},  the upper bound for bufferless $\cdot$/M/1 queue, and the capacity $R(\lambda,\mu)$ for the M/M/1 bufferless queue. All the systems have exponential service time of rate $\mu$ and arrival  rate $\lambda$. Both BTQ and bufferless upper bound plots have maximum equal to 0.3679 nats per average service time whereas the maximum achievable rate is 0.3340 nats per average service time.}
\end{figure}

Fig.~\ref{fig:Upperbound-Comp} illustrates the achievable upper bound of M/M/1 (\ref{eqn:Achievab_MM1}) and the universal upper bound of $\cdot$/M/1   (\ref{eqn:MM1Upper}). The $\cdot$/M/1 upper bound of "Bits through queues" (BTQ) paper \cite[eq.~2.17-2.18]{anantharam1996bits} is plotted for comparison. In these plots, $0<\lambda/\mu<1$ since the $\cdot$/M/1 BTQ requires $\lambda\leq \mu$ for stability.  From this plot, we can see that the maximum value of the upper bound of $\cdot$/M/1  is equal to the maximum value of $\cdot$/M/1 BTQ which is 0.3679 nats per average service time.

\subsection{Achievability for the M/G/1 queue}
\begin{thm}\label{thm:Achievability-MG1}
The M/G/1 bufferless queue $S$ with arrival rate $\lambda$ and average service time $\E{S}=1/\mu$ has capacity
\[
C(\lambda,S)\geq R(\lambda,\mu).\]
\end{thm}
\begin{IEEEproof}
The procedure for this proof is along the lines of the proof of \cite[Thm.~7]{anantharam1996bits}.
We assume the inter-departure times under general service have PDF $\fg[\dseq]{\Dseq}$, and the arrivals 
are a rate $\lambda$ Poisson process.
We further assume
that $\fe{\Dseq}$ is the PDF of the inter-departure times of system with a memoryless server of rate $\mu$ (which would be hypoexponential). 
Now similar to \cite{anantharam1996bits},
\begin{align}
 & i_{\Aseq;\Dseq}(\Aseq;\Dseq)\nonumber\\
 & \qquad=\log\frac{\fg{\Dseq|\Aseq}}{\fg{\Dseq}}\eqnlabel{InfDenGen_1}\\
 & \qquad=\log\frac{\fg{\Dseq|\Aseq}}{\fe{\Dseq|\Aseq}}
 -\log\frac{\fg{\Dseq}}{\fe{\Dseq}}
 +\log\frac{\fe{\Dseq|\Aseq}}{\fe{\Dseq}}.
 \eqnlabel{InfDenGen_2}
\end{align}
In Theorem~\ref{thm:Achievab_MM1}, we showed that
\begin{align}
\liminfp \frac{1}{T_n}\log\frac{\fe{\Dseq|\Aseq}}{\fe{\Dseq}}
 =R(\lambda,\mu).\eqnlabel{InfDenGen_4}
\end{align}
We need to prove that 
\begin{align*}
\liminfp \frac{1}{T_n}\left[\log\frac{\fg{\Dseq|\Aseq}}{\fe{\Dseq|\Aseq}}-\log\frac{\fg{\Dseq}}{\fe{\Dseq}}\right] \ge 0.
\end{align*}
Note that \eqnref{Time-iid} implies it is sufficient to prove that for every $\zeta>0$,
\begin{align*}
\lim_{n\rightarrow\infty}
\prob{\frac{1}{n}\left(\log\frac{\fg{\Dseq|\Aseq}}{\fe{\Dseq|\Aseq}}-\log\frac{\fg{\Dseq}}{\fe{\Dseq}}\right)\leq-\zeta}= 0.
\end{align*}
 Using the same method as \cite{anantharam1996bits},
\begin{align}
 & \prob{\frac{1}{n}\left(\log\frac{\fg{\Dseq|\Aseq}}{\fe{\Dseq|\Aseq}}-\log\frac{\fg{\Dseq}}{\fe{\Dseq}}\right)\leq-\zeta}\nonumber\\ 
 &=\prob[\fg{\Aseq,\Dseq}]{\frac{1}{n}\log\frac{\fg{\Dseq|\Aseq}}{\fe{\Dseq|\Aseq}}\frac{f_{\Aseq}f_{\Dseq}}{\fe{\Aseq}\fg{\Dseq}}\leq-\zeta}\\
 & =\prob[\fg{\Aseq,\Dseq}]{%
 \frac{1}{n}\log\frac{\fg{\Aseq|\Dseq}}{\fe{\Aseq|\Dseq}}
 \leq-\zeta}\\
 & =\iintop_{\fg{\Aseq|\Dseq}\leq e^{-\zeta n}\fe{\Aseq|\Dseq}}\hspace{-1.2cm}\fg[\xseq|\yseq]{\Aseq|\Dseq}
 \fe[\yseq]{\Dseq}d\xseq\,d\yseq\\
 & \leq e^{-\zeta n}\iint \fe[\xseq|\yseq]{\Aseq|\Dseq}\fe[\yseq]{\Dseq}d\xseq\,d\yseq\\
 & =e^{-\zeta n}.\eqnlabel{NonnegativeSubtract}
\end{align}
It follows that 
\begin{equation}
\liminf_{n\rightarrow\infty}-\mathrm{p}\frac{1}{T_n}\left(\log\frac{\fg{\Dseq|\Aseq}}{\fe{\Dseq|\Aseq}}-\log\frac{\fg{\Dseq}}{\fe{\Dseq}}\right)\geq0.\eqnlabel{Nonnegative-liminf}
\end{equation}
Now using (\ref{eqn:InfDenGen_4}) and (\ref{eqn:Nonnegative-liminf}),
we see that for every $\zeta'>0$,
\begin{align*}
\lim_{n\rightarrow\infty}P\left[\frac{1}{T_{n}}i_{\Aseq;\Dseq}(\Aseq;\Dseq)\leq R(\lambda,\mu)-\zeta'\right]= 0.
\end{align*}
Thus Theorem~\ref{thm:Achievability-MG1} holds.
\end{IEEEproof}
It must be noted that this is not necessarily a tight lower bound
similar to \cite{anantharam1996bits}.
 The result of Theorem~\ref{thm:Achievability-MG1} shows that the exponential server has the lowest capacity for a fixed service rate among servers with Poisson arrivals.

\section{conclusion\label{sec:conclusion} }

This paper studied the capacity of timing channels described by bufferless
single-server timing queues with iid service times. One of the main
challenges in the analysis of such timing channels is the lack of a
one-to-one correspondence between packets arriving at and departing from the
queue. This challenge was circumvented by resorting to codewords with
infinite length, with the rate of the code defined using the average time it
takes to observe the departure of $n$ codeword packets. In general, we
believe that an information-theoretic understanding of the setup studied in
here will help us address the challenge of causal inference in systems, such
as (online) social networks, that lack a one-to-one correspondence between
different actions (e.g., tweets versus retweets). In this regard, this paper
discussed the maximum likelihood decoder for decoding timing messages
transmitted through a bufferless queue, provided upper bounds on the channel
capacity---including a single-letter upper bound and a looser universal upper
bound, and computed achievable rates for bufferless M/M/1 and M/G/1 queues.
Computing tighter upper bounds on the capacity and achievable rates for
$\cdot$/M/1 and $\cdot$/G/1 queues that meet the upper bounds remain areas of
future work.

\appendix{ \label{appendix:a}
In this part, the upper bound (\ref{eqn:MM1Upper}) for the M/M/1 queue  is shown to be only a function of $\rho=\lambda/\mu$ for fixed $\mu$. Initially, the $h_{\text{hypo}}(\lambda,\mu)$ is rewritten using \eqnref{hypo-pdf} as follows:
\begin{align}
 & h_{\text{hypo}}(\lambda,\mu)\\
 & =-\intop f_{D}(x)\log f_{D}(x)dx\\
 & =-\intop f_{D}(x)\log\left[\mu e^{-\mu x}\frac{\rho}{1-\rho}\left(e^{-\left(\lambda-\mu\right)x}-1\right)\right]dx\\
 & =-\log\mu+\mu E\left[D\right]\log e-\log\left(\frac{\rho}{1-\rho}\right)\nonumber \\
 & \qquad-\intop f_{D}(x)\log\left(e^{-\left(\lambda-\mu\right)x}-1\right)dx.
\end{align}
With the change of variable $y=\left(\lambda-\mu\right)x$,
\begin{align}
h_{\text{hypo}}(\lambda,\mu) & =-\log\mu+\left(1+\frac{1}{\rho}\right)\log e\nonumber \\
 & \qquad -\log\left(\frac{\rho}{1-\rho}\right)+\frac{\rho}{\left(1-\rho\right)^{2}}G\left(\rho\right),\eqnlabel{RewrittenUpper}
\end{align}
where 
\[
G\left(\rho\right)=\intop e^{-\frac{y}{\rho-1}}\left(e^{-y}-1\right)\log\left(e^{-y}-1\right)dy
\]
 is a function of $\rho$. Now substituting (\ref{eqn:RewrittenUpper})
in \eqnref{Rdefn}, 
\begin{align*}
R(\mu,\lambda)& =\mu\frac{\log e
-\rho\log\left(\frac{\rho}{1-\rho}\right)+\left(\frac{\rho}{1-\rho}\right)^2 G\left(\rho\right)}{1+\rho}
\end{align*}
which proves the claim.}
\bibliographystyle{IEEEtran}
\bibliography{MyDatabase}

\end{document}